# ICOA638: ANALYSIS OF A SITE'S INTEGRITY BY 3D MODELS AND INTEGRATED DATABASE, CASE STUDY: THE PIC-DU-MIDI HIGH-MOUNTAIN OBSERVATORY (FRANCE)

**Subtheme 03: Protecting and Interpreting Cultural Heritage in the Age of Digital Empowerment**

**Author: Michel Cotte, Florent Laroche, Matthieu Quantin, Loïc Jeanson, Nicolas Bourgeois**

*Michel Cotte is emeritus professor of the History of Technology; he is also an expert and adviser for ICOMOS. He was editor for ICOMOS of thematic study on the Heritage of Astronomy (2010) and Cultural Heritage of Water Management, case studies of Middle East and Maghreb (2014).*

**Abstract:** The concept of "integrity", as currently used in the analysis of World Heritage sites or cultural landscapes mainly consists of 1) the composition of a given site, regarding its origins and its current state of conservation; 2) the visual and functional relationships between its components (attributes).

One of the major questions is "what defines the origin period?". The integrity analysis has to clearly understand and evaluate which tangible components exist, in order to correctly identify their origin period, estimate how much of the original structure and function remain.

An additional difficulty rises in case of scientific and/or technical heritage assessment: the very historical necessity of updating and implementing technical/scientific innovations. These follow advances in science and/or technique(s), and lead to frequent successive changes, impacting the site's structure. Hence, for living sites, one cannot see the origin of the project as the only reference state, as it could be for classical heritage; it is instead requisite to enlighten a chronological series of major reference states.

The Pic-du-Midi Observatory (France) is a remarkable example as high-mountain scientific station (2860 m). Its various and successive scientific uses led to a series of reshaping and structural evolutions over around 150 years, till nowadays. We have there a series of reference states very conform to an active scientific station and observatory.

The project aims to combine historical data and measurements on the current site's state, to build a succession of 3D models for a series of 5 or 6 'reference state' of the observatory within associated database, and to document the integrity analysis more accurately than usual. The final uses of the digital achievement will also serve for the management of the site and for further development of pedagogical historical movies for visitors (Pic-du-Midi houses a museum of astronomy and physical sciences).

ICOA638: ANALYSIS OF A SITE'S INTEGRITY BY 3D MODELS AND INTEGRATED DATABASE, CASE STUDY: THE PIC-DU-MIDI HIGH-MOUNTAIN OBSERVATORY (FRANCE)

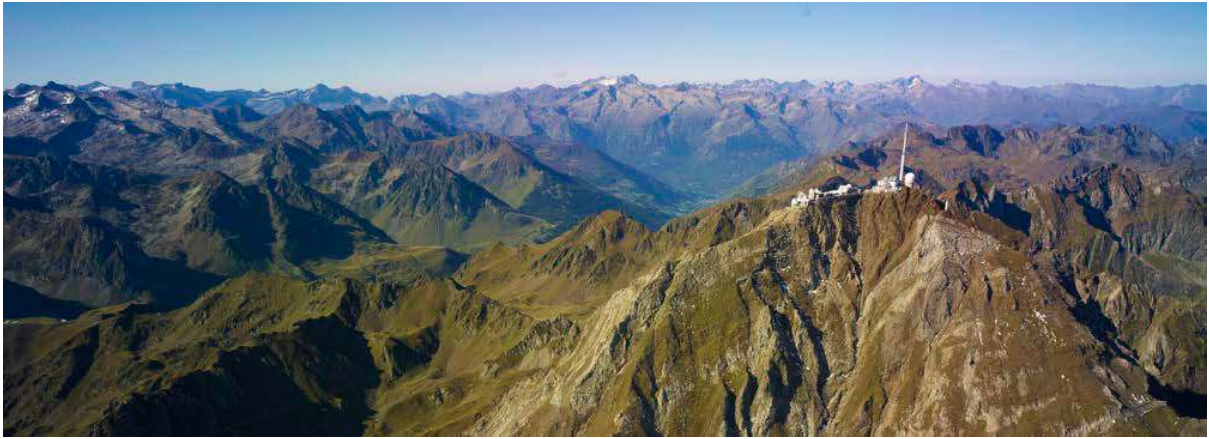

*Fig.: Pic-du-Midi Observatory and Scientific Station: in regular use from 1883[1]  © Régie Pic-du-Midi*

## Introduction: terms of reference for integrity

ICOMOS usual practice
The concept of "integrity" is currently used today for the analysis of World Heritage cultural sites or cultural landscapes, as it was from a longer time for natural heritage. To be conformed to the format, every dossier must contain an integrity assessment. The major methodological trend observed by ICOMOS evaluation focuses on three main complementary facets, implying different scales of analysis:

1) <u>Inventory level</u> of a given site: the attributes must have sufficient number, readability and meanings; the set completion and its wholeness must be remarkable, exceptional or even unique, regarding its origins and its current state of conservation; this analysis level is fundamental and can't be bypassed;

2) <u>Landscape level</u>: the visual and structural relationships between attributes (both cultural and natural) must be remarkable, exceptional or even unique, making up the panorama quality of the site and expressing its value in a comprehensive way; this point is very important and in case of insufficient quality, visual aggression, lake of appropriate regulation (buffer zone) it could compromise the recognition;

3) <u>Intangible level</u>: integrity of use involving concepts of "living place" or "evolving landscape"; it must be maintained with remarkable, exceptional or even unique meanings and historical significances; the use/usage and its continuity is something very important, and it will be certainly upgraded in close future as requirement for modern and contemporary heritage, especially in the field of science and technology.

---

[1] https://www3.astronomicalheritage.net/index.php/show-entity?identity=76&idsubentity=1

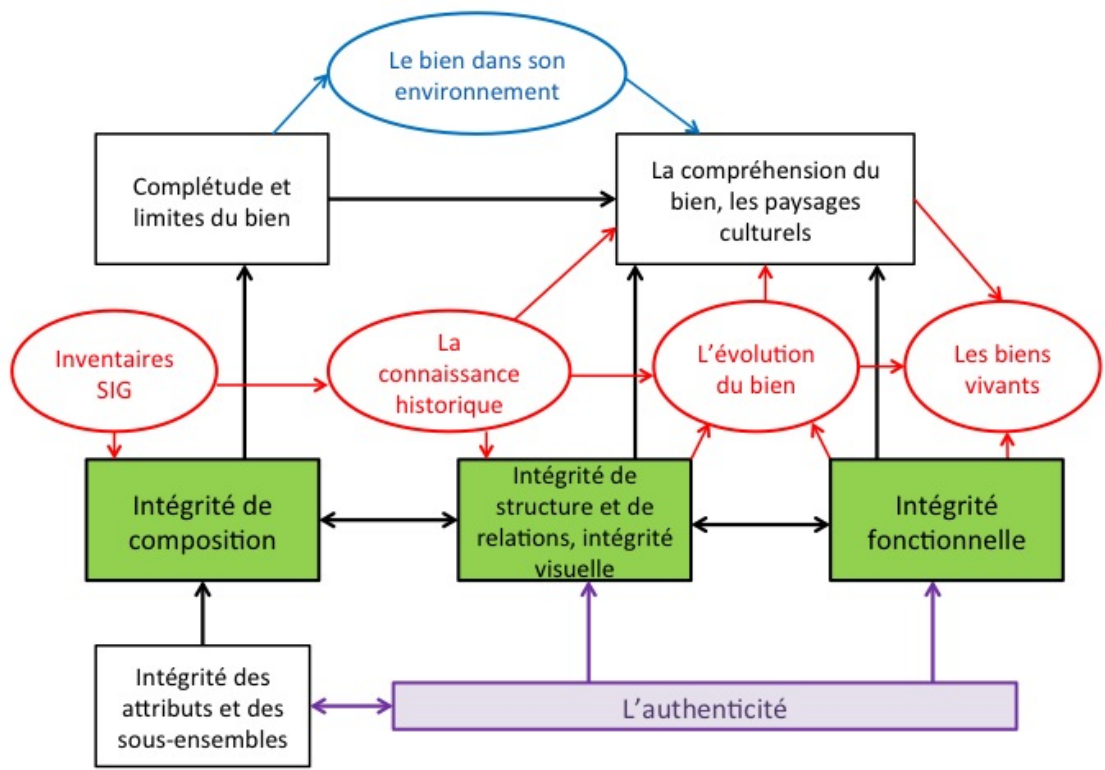

Operational Guidelines of the WH Convention

On the other hand, the *Operational Guidelines for the implementation of the World Heritage Convention* gives us the layout for the definition of integrity and about the way for using it for nominations, in a very broad sense both for cultural and natural heritage:

> **§.88** *Integrity is a measure of the wholeness and intactness of the natural and/or cultural heritage and its attributes. Examining the conditions of integrity, therefore requires assessing the extent to which the property:*
> - *a) includes all elements necessary to express its Outstanding Universal Value;*
> - *b) is of adequate size to ensure the complete representation of the features and processes which convey the property's significance;*
> - *c) suffers from adverse effects of development and/or neglect.*
>
> *This should be presented in a statement of integrity.*
>
> **§.89** *For properties nominated under criteria (i) to (vi), the physical fabric of the property and/or its significant features should be in good condition, and the impact of deterioration processes controlled. A significant proportion of the elements necessary to convey the totality of the value conveyed by the property should be included. Relationships and dynamic functions present in cultural landscapes, historic towns or other living properties essential to their distinctive character should also be maintained.*

There is more important recommendations and detailed issues about natural properties under criteria (vii) to (x), but an interesting remark is still there in the July 2017 last edition:

> *Examples of the application of the conditions of integrity to properties nominated under criteria (i) - (vi) are under development.*

These reference points give us the background of our researches, and we are on the way to propose some case study, developments and epistemological remarks. The pivotal issue of this communication is: "how to apply the integrity concept in the case of an evolving place, especially a cultural landscape with major scientific meanings and constraints?" Our purpose is to study a case of scientific living site and to define its integrity using digital tools issued from engineering sciences.

**Case study of Pic-du-Midi: a living high mountain scientific site**

An epistemological question for scientific and technical heritage

One of the major questions for a technical/ scientific heritage site is: "what could define the origin period?" We clearly face an important difficulty/ impossibility and every case must be carefully assessed in this way because many living places in past times and perhaps till today have frequent evolutions, and even permanent evolution. We are in front of a kind of contradiction/ opposition between the classical architectural point of view of style period and its aesthetic expression, and of scientific heritage as a living place in permanent renewal, both for scientific programs and for techniques/ instruments. Architectural change has a very different meaning from classical point of view. There it is like "collateral effect" of deeper changes; it is a subsequent consequence.

So, the integrity analysis has to clearly understand and evaluate such succession of situation and which tangible components exist now, in order to correctly identify their implementation period, to estimate how much of the original structure and function remains today. That must be done in a comprehensive way for scientific context.

A pivotal difficulty rises in case of scientific and/or technical heritage assessment: the very historical necessity of updating and implementing technical/ scientific innovations. These follow advances in science and/or technique(s), and lead to frequent successive changes, impacting the site's structure. Hence, for living sites, one cannot see the origin of the project as the only reference state, as it could be for classical heritage; it is instead requisite to enlighten a chronological series of major reference states. Such specificity of scientific/ technical places leads us to a new concept of succession of identified reference state without any *a priori* hierarchy between them. We have now to present a case study:

Rapid historical overview

The Pic-du-Midi Observatory (France) offers a remarkable example as high-mountain scientific station (2860 m). Its various and successive scientific uses led to a series of reshaping and structural evolutions lasting all along around 150 years of scientific history, till nowadays where it remains an active professional observatory (University of Toulouse) and a famous high mountain tourism site.

All along its story, it was a scientific station in a large sense even it was mainly seen as an astronomical observatory today and from the middle of 20th Century. Indeed astronomical observations were done frequently all along its different historical periods. At its origins (years 1870-80) it was mainly a meteorological and atmospheric station studying for instance clouds, atmosphere lightning, rain regime, ozone; along with altitude botanic and biology, human behaviour in wintering, etc. After that and in parallel to the astronomical observatory project (from 1883), a series of projects with appropriate constructions and scientific implementations have happened, until nowadays projects of adaptation of astronomical observatory by the university and touristic program (museum, tourism, skiing, etc.).

The almost permanent construction/ evolution of the station was mainly guided by a series of important successive and recurrent facts: need of permanent human presence, even during winter (uninterrupted scientific wintering since 1884 – even during war periods), lack of accessibility (road construction is impossible), limited available surface on the Pic and geological rocky context in altitude for civil engineering, human needs (food, energy, telecommunication) and technical/ scientific instruments.

A succession of reference states all along 150 years of scientific use

Consequently, we have identified a series of reference states very conform to an active scientific station and observatory, and the first question for evaluation of the integrity question is to perform a precise inventory in a couple of ways:
1) What was really done at each reference period of the history of the Pic-du-Midi? and
2) What remains today, as visible or not-visible components of the site?

Site topology and today ensemble of buildings, access equipment, facilities, scientific and technical components, and etc. made of a complex set of elements, which from original construction or as later modified construction. Imbrication of attributes is complex, therefore so is inventory and even more complex is the analysis of integrity and authenticity in the way of upper point 1): inventory level and assessment of the completeness. An additional challenge is to present and to explain this complexity to visitor.

The point 2 of the integrity current assessment practice for an evolving landscape proceeds directly from the 3D compilation associated with iconic database within appropriate information. It shows exactly what means an evolving and living landscape. In this case study, point 3 of the continuity of use is obvious by the history of recent development and future projects.

Fortunately, documentation is consequent even it is not complete and scattered between numerous public and private owners. There is especially a large series of photos, count of the exceptionality both of the project at different times and of the panorama associating human works and spectacular high mountain pic. Finally, we selected six main successive states of reference, but really speaking it is almost a continuous and permanent maintenance and refurbishing of the place!

First part of the work was done for inventory of archives and digital copy of documentation. We also had a physical scale model of the Pic-du-Midi, today used as a piece of the museum exhibition. It was done with accurate attention and high professional knowhow, which support notable information about tangible evidences.

**The use of a couple of correlated digital tools for integrity analysis**

The project aims first to combine historical data and today measurements to build a succession of 3D models for a chronological restitution of the history of construction by a series of 6/7 successive slides of 'reference states' of the observatory. These virtual reconstructions must be 'scientific' in the way to be closer as possible from historical geometric truth, aesthetic consideration come in second.

Next and also, the project aims to associate a database (in some way 'intelligent database') with the 3D models, and to document the integrity analysis more accurately and more quickly than usual. For that we hope to correlate database and 3D model and to open field of applications in different ways.

Sources

The technical difficulty for correctly mastering data and producing valuable 3D successive representations relies on different resources, but with some inherent weaknesses:
- We have the privilege to meet a significant amount of archives and photos, from different periods, and also some valuable historical studies among them 'Davoust publication' is certainly the most complete[2]. Nevertheless, these archives resources are frequently scattered in different independent public and private collections; on the other hand, these documents are not exhaustive with some missing information especially about construction or from a not registered origin and/or time; images are frequently repetitive and mainly done for visitor as 'postcard';
- Photos are relatively abundant and of quality especially for some ancient 'black and white' series; but generally speaking, photos are related to two problems: 1) lack of date for many, and 2) reference views came from the same and unique geographical point: the summit of the Pic, because it offers a close and remarkable eye-bird view. The other viewpoints on the Pic are missing for obvious reason due to the shape of the Pic, and we have to wait for aerial photos with enough precision to recent period.

The 3D scale chronological model with 'time slider'

For the first phase, we tried to do successive digital clouds in chronological order. That aimed to present a scale-evolving model driven directly by a button screen slider. It involves the idea of successive state of reference, using colours. It also used naming of buildings.

We use the following document and methodology for doing digital clouds:
- Attempts of photogrammetry on historical photos (not precise and mainly from one point of view as stated upper);
- Scans of a physical scale model is sued from the 1980s (excellent quality of geometrical replica checked with 2017 aerial scan of the mountain);
- Recent aerial photos for gathering different external and missing viewpoints,
- Some limited and experimental attempts for manual scanning on today external constructions.

So digital works for making one 3D model from the different clouds relied on following main issues:
- To gather and to harmonise data information for having an unique and suitable digital cloud norm and scale;
- To assess common geometrical references and to paste together the different clouds representing different section of the Pic structure;
- To build the successive reference state of the Pic-du-Midi, with temporal function allowing the user to slide from one period to another;
- To manage colours and contrasts with one trigger for user;

In methodological and ethic terms, we chose with absolute priority for the restitution of geometrical volumes with a maximum of conformity to historical truth. We refused to make aesthetic corrections or hypothesis to complete missing parts. We are aware that the result seems very rough sometime and not very refined, but it aims to be more of a reflection of historical reality than a pleasing to the eye interpretation. It bases the integrity reliance even it is incomplete and not very refined. An associated problem rise: is it possible to estimate margin of error?

---

[2]DAVOUST Emmanuel, 2000, *L'Observatoire du Pic-du-Midi, Cent ans de vie et de science en haute montagne*, Paris : CNRS Editions.

The associated database and correlation with 3D scale model

Classical database will be the first step of our implementation, by an ordered compilation of text archives, photos and publications within indexation, keywords, research by simple requests as: name of building, date, actors and etc. It is currently under process and he helped strongly to build and to publish the present chronological 3D models. It could work as reference tool for each building, giving a set of information: location and architectural form, date of implementation, initial purpose of use, name of designer and sometime of builder, possible evolution and dates of evolution. In association with the model under construction, it helps us to understand the way of design of the individual buildings and facilities, and of the ensemble. It allows to understand conception and use of limited surfaces and volumes in drastic winter conditions for human life and scientific regular practice (more than 2800 m high in European climate context).

The second purpose is to deepen database functionalities and give it possibility of pre-analytic contents in addition to basic answer. For that we examine to implement a specific internal architecture allowing database to be 'intelligent' and 'cooperative' in real-time of research done by the couple historian - model-builder. It seems possible to propose to them additional information and suggestions when data base present the basic responses to a request; e.g. possible complementary information, neighbour fields, association with other fields or persons, and etc. For that we are paying attention about some fields of researches as:
- Textual analysis of printed document at basic and medium level;
- Image analysis of photos at basic level;
- The question of associations by the graph method.

Finally, the purpose of inter-operability with 3D model is also examined, count of an important former experience with researches par the Nantes scientific team in association with the City of Nantes Museum (Castel of Brittany Dukes) on the project "Harbour of Nantes 1900"[3].

**Conclusion and perspectives**

Pic-du-Midi

The initial purpose of the project is clearly to build a dynamic association of 3D models in chronological order and database to document accurately and surely the question of inventory of scientific station implying important and regular historical evolutions. Pic-du-Midi is obviously a wonderful example of such situation, both because its specific constraints due to the goal of a permanent use of high mountain summit and by an exceptional ancient and regular use and renew of scientific programs from around one century and half.

Such 'integrity of composition' study supposes a large and precise knowledge to deliver and to document a complex story of construction at the level of individual buildings and scientific rooms. 3D models give us an image with scientific certitudes about historical existence of a given artefact and checking of its today evidence. To complete the process done at this step we intend to launch a second campaign of geometric documentation, by individual scanning of the present situation of every construction at Pic-du-

---
[3] http://www.club-innovation-culture.fr/chateau-de-nantes-maquette-nantes-1900-ecrans-tactiles-crowdsourcing/

Midi, outside and inside. Confrontation of the two approaches should give us final conclusive results for the integrity analysis, and to refine visual quality of final 3D model.

Beyond this very specialised approach, we also intend to deepen the 3D model with description of the inside facet of scientific rooms and main living rooms, and further the restitution of the most important scientific instruments of World importance for the history of science in context of use; e.g.: the Bernard Lyot Coronagraph, the Wilson Chamber for discovery of the Hyperon elementary particle, and etc.
The final uses of the digital achievements will also serve as tool for the management of the site and for further development of pedagogical historical movies for visitors. The Pic-du-Midi houses a museum of astronomy and physical sciences, perhaps the highest in the World!

## Theoretical perspective of the ReSeed research project

Nowadays, the situation of technical and industrial heritage poses many problems: how to manage and valorise it in the framework of museums and on sites? How to prolong the longevity of the technical information of these collections, archives and heritage sites? This technical information, a testimony of the past, is becoming old very quickly and not very readable by new generations; like a puzzle, in which the tangible parts and intangible ones wear out or disappear, the technical data disperses progressively with time. That is why preserving the national technical patrimony has become a priority for governments and organisations worldwide. Since keeping and maintaining physical objects in museums is expensive and very frequently purely static; so our approach proposes a new way of conservation that preserves these articles as numerical objects and recreate a renewal of integrity and dynamics by vitality.
Reverse engineering is widely used in manufacturing industries in order to record geometrical data from the products. However, digitizing physical data does not provide any knowledge about the products. Capitalizing this know-how and knowledge is critical to advance tomorrow's products. So we are pursuing two complementary approaches as seen for Pic-du-Midi example:
– Classical knowledge management based on semantic sources,
– Computer assisted design and scanning tools.

Yet, there still does not exist any global process: technical issues born from interdisciplinary processes drive us to scientific issues: there is an important need to bring the different models into line and to carry out a case-by-case review of the usual industrial processes.

The French national project 'ReSeed' (www.reseed.fr) aims for the development of a new technology: a tool and an interoperable format in order to digitize both semantic and physical data about objects. In the end, in case of success of the ANR ReSeed program, a cycle between the past and the present knowledge will be established. Actually, protecting, analysing and understanding these objects from the "past heritage" may turn them into "present capital". The transposition of past knowledge into contemporary form, readable and intelligible in our present socio-technical system, may become a source of inspiration to anticipate our future.

The tool developed during the ReSeed project offers multi-level access in order to capitalize and use a digital heritage object for helping to demonstrate the Heritage Value.

**Authors and institutions:**

---

[4] http://www.agence-nationale-recherche.fr/Projet-ANR-16-CE38-0005,
www.reseed.fr,
[5] http://www.obs-mip.fr/pic-du-midi,

# ICOA638: ANALYSE DE L'INTÉGRITÉ D'UN SITE PAR LA MODÉLISATION 3D AVEC UNE BASE DE DONNÉES INTÉGRÉE, ÉTUDE DE CAS: LE PIC-DU-MIDI, OBSERVATOIRE DE HAUTE MOTAGNE (FRANCE)

**Sous-thème 03: Protéger et interpréter le patrimoine culturel à l'ère de l'autonomisation numérique**

**Session 1:** Pertinence des outils numériques et de la technologie dans la documentation, la conservation et la sauvegarde du patrimoine et l'engagement communautaire
**Lieu:** Silver Oak 2, India Habitat Centre
**Date et heure:** 13 Décembre, 2017, 09:10 – 09:25

**Auteur: Michel Cotte, Florent Laroche, Matthieu Quantin, Loïc Jeanson, Nicolas Bourgeois**

Michel COTTE est professeur émérite d'Histoire de la Technologie; il est aussi expert et conseiller pour ICOMOS. Il a été éditeur des études thématiques sur le patrimoine de l'astronomie (2010) et sur le patrimoine culturel de la gestion de l'eau, avec des études de cas au Moyen-Orient et au Maghreb (2014)

**Abstract:** Le concept d '«intégrité», tel qu'il est actuellement utilisé dans l'analyse des sites du patrimoine mondial ou des paysages culturels, consiste principalement en: 1) la composition d'un site donné, notamment ses origines et son état actuel de conservation; 2) les relations visuelles et fonctionnelles entre ses composants (attributs).

L'une des principales questions est "qu'est-ce qui définit la période d'origine?". L'analyse d'intégrité doit clairement comprendre et évaluer quelles composantes tangibles existent, afin d'identifier correctement leur période d'origine, et d'estimer combien il reste de la structure et de la fonction d'origine.

Une difficulté supplémentaire se pose dans les cas d'évaluation du patrimoine scientifique et/ou technique: la nécessité historique de suivre et de mettre en œuvre les innovations techniques/scientifiques. Celles-ci suivent les progrès de la science et/ou de la (des) technique (s), et conduisent à des changements fréquents et réguliers, qui modifient la structure du site. Par conséquent, pour les sites vivants, on ne peut pas se limiter à l'origine du projet comme étant le seul état de référence, comme cela peut être le cas pour le patrimoine classique, mais on doit plutôt mettre en évidence une suite chronologique des principaux états de référence.

L'Observatoire du Pic-du-Midi (France) est un exemple remarquable de station scientifique de haute montagne (2860 m). Ses diverses utilisations scientifiques successives ont conduit à une série de remodelage et d'évolutions structurelles sur une période d'environ 150 ans, jusqu'à nos jours. Nous avons là une série d'états de référence très conformes à une station scientifique active et à un observatoire.

Le projet vise à combiner des données historiques et des mesures sur l'état actuel du site, afin de construire un ensemble de modèles 3D pour une série de 5 ou 6 « états de référence » de l'observatoire dans une base de données associée dans le but de documenter plus finement l'analyse d'intégrité. Les produits finaux de cette réalisation numérique serviront également à la gestion du site et au développement de films historiques pédagogiques pour les visiteurs (le Pic-du-Midi abrite un musée de l'astronomie et des sciences physiques).

*Mots-clés:* *intégrité, modèle numérique, base de données, observatoire*